# Calculated Effects of Vacancy and Ti-doping in 2D Janus MoSSe for Photocatalysis

Yi-Ming Zhao[1], Pengju Ren[2], Xing-Yu Ma[3], James P. Lewis*[2,4], Qing-Bo Yan*[1]

and Gang Su*[5,3,1]

[1]Center of Materials Science and Optoelectronics Engineering, College of Materials Science and Opto-electronic Technology, University of Chinese Academy of Sciences, Beijing 100049, China.

[2]State Key Laboratory of Coal Conversion, Institute of Coal Chemistry, Chinese Academy of Sciences, Taiyuan, Shanxi 030001, China.

[3]School of Physical Sciences, University of Chinese Academy of Sciences, Beijing 100049, China.

[4]Beijing Advanced Innovation Center for Materials Genome Engineering, Beijing Information S&T University, Beijing 101400, China.

[5]Kavli Institute for Theoretical Sciences, and CAS Center of Excellence in Topological Quantum Computation, University of Chinese Academy of Sciences, Beijing 100190, China.

# ABSTRACT

Two-dimensional (2D) Janus transitional metal dichalcogenides (TMDCs) have great potential for photocatalytic water splitting due to their novel properties induced by the unique out-of-plane asymmetric structures. Here, we systematically investigate the geometric, electronic and optical properties of 2D Janus MoSSe with titanium doping and vacancies to explore their synergistic effects on photocatalytic activity. We find that there are effective attractions between the substituted or adsorbed Ti atoms and S/Se vacancies. The Ti adatoms dramatically extend the light absorption range to infrared region. The S/Se vacancies coexisting with Ti adatoms will modulate the transition of photo-excited electrons, thereby enhancing the sunlight absorption. The Ti adatoms either existing alone or coexisting with vacancies introduce smaller lattice distortion compared to substituted Ti atoms and these Ti adatoms induce smaller effective mass of charge carriers. The configuration of S vacancy coexisting with Ti adatoms on Se-surface exhibits the most significant synergistic effects and best overall photocatalytic performance. Our work reveals the mechanism and effects induced by doping and vacancies coexisting in 2D Janus TMDCs, also propose a new practical strategy to improve the performance of 2D photocatalysts.

## 1. Introduction

Photocatalytic water splitting is a clean and renewable way to reduce the consumption of carbon-based fossil fuel and mitigate the greenhouse effect.[1] Photocatalysts such as $TiO_2$,[2, 3] CdS,[4] $Fe_2O_3$,[5, 6] and TaON,[7] etc. can catalyze redox reactions of water splitting under sunlight; however, limited performance restricts broader applications.[8] The water splitting reaction contains the hydrogen evolution reaction (HER) and oxygen evolution reaction (OER) as Figure 1 shows. HER referring to the reduction of $H^+$ ions to $H_2$ molecule is a crucial step for the $H_2$ generation via water splitting. The best performing photocatalysts should have proper band-edge energy alignments, significant carrier mobility, suitable bandgaps, and proper adsorption of surface reactants.[9] The valence/conduction bands should straddle the redox reaction potentials to enable the photo-generated electrons and holes to transfer to these potentials and catalyze the redox reaction.[10] The carrier mobility determines the efficient separation of electron and hole pairs. A suitable bandgap enables the semiconductors to effectively harvest sunlight energy. Finally, the proper adsorption of surface reactants will improve the efficiency of surface reactions. Two-dimensional (2D) semiconductors are promising high-performance photocatalysts due to a larger specific area for surface adsorbates to react and shorter distances for charge carriers to migrate to the surface.[11, 12]

Among 2D semiconductors, monolayer molybdenum disulphide ($MoS_2$) exhibits potential as a material for photovoltaic and photocatalytic applications because of its proper band edges for catalyzing the water splitting reaction,[13, 14] greater mobility of charge carriers compared to many semiconductors,[15] and a direct bandgap of 1.73 eV which is in the visible range.[14, 16] However, perfect $MoS_2$ has disadvantages such as lacking catalytically-active sites due to its very smooth surface morphology. Additionally, the fast recombination of photogenerated electron-hole pairs, while excellent for photoluminescence, diminishes opportunities for the absorption energy of MoS$_2$ to be utilized in chemical reactions.[17, 18] Recently, Lu *et al.* synthesized a novel 2D Janus-structured MoSSe (2D J-MoSSe) successfully by completely replacing all sulfur atoms on one surface of the 1H-phase MoS$_2$ monolayer with selenium atoms via chemical vapor deposition (CVD).[19] The 2D J-MoSSe obtains breaking out-of-plane symmetry as differences between S and Se, such as electronegativity and atomic radius, *etc*. This breaking out-of-plane symmetry induces an intrinsic internal electric field in the 2D J-MoSSe, pointing from the Se plane to the S plane and perpendicular to the surface.[20] There is a positive and negative charge accumulation around the Se and S atoms, respectively, resulting from the larger electronegativity of sulfur compared to selenium.[21] The internal electric field will boost the separation of charge carriers as well. A direct bandgap of 1.68 eV enables 2D J-MoSSe a slightly better utilization of sunlight compared with MoS$_2$.[22] Guan *et al.* hypothesized, from DFT calculations of optical-absorption spectra and band-edge alignments, that the 2D J-MoSSe may have significant photocatalytic performance.[22] Figure 1 illustrates how band edges are modified by the electrostatic potential difference ($\Delta V$) caused by the changes in morphology. The Se and S sites attract photo-generated electrons and holes, respectively, and the band edges bend upward along the direction of internal electric field.[23, 24] The electrons in the conduction band (localized on the Se surface) increase in energy due to this band-edge shift of $\Delta V$. The higher-energy electrons will more readily transport to the reduction potential thereby more readily reducing the H$^+$ ions to H$_2$ while the holes more readily transport to the oxidation potential thereby oxidizing the H$_2$O molecule to O$_2$.

The ideal catalyst should adsorb the protons within the right window of binding energies (neither too weakly nor too strongly) to catalyze the HER efficiently (for a more thorough discussion of HER see the Supplementary Information). Unfortunately, the basal planes of pristine 2D J-MoSSe have difficulty adsorbing reactants (protons and water molecules).[20] The intrinsic S or Se vacancies will adsorb protons better than the basal planes thereby significantly improving the efficiency of HER on the surfaces of 2D J-MoSSe, as shown by both experiment and theoretical results.[20, 25, 26] The S or Se vacancies naturally exist on the surfaces of 2D J-MoSSe when prepared by CVD, similar to $MoS_2$ which contains a noticeable fraction of S vacancies when prepared by either mechanical exfoliation or CVD.[27, 28]

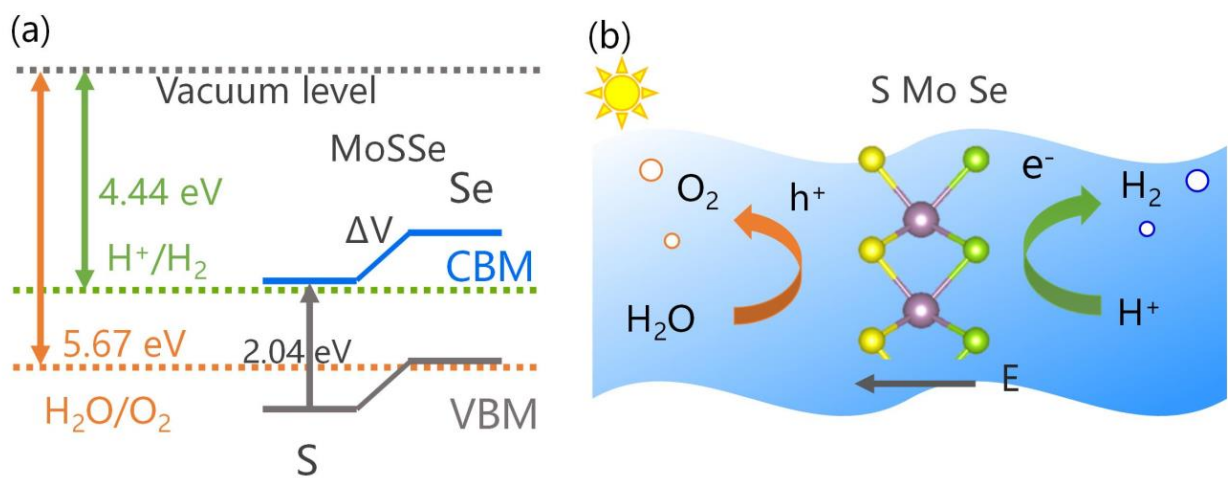

**Figure 1.** (a) Schematic illustration of the band edges of 2D J-MoSSe with respect to the redox potential of water splitting reaction and the bending of energy levels caused by the internal electric field. (b) The distribution of photogenerated electrons and holes and the schematic diagram of HER and OER reaction.

Researchers also demonstrated that adsorbed transition-metals on 2D J-MoSSe will improve the reactant adsorption as well as enlarge the optical absorption range.[29] Transition metals adsorbed on the Se (S) surface of 2D J-MoSSe will strengthen (weaken) the internal electric field.[21] Intrinsic S or Se vacancies will exist simultaneously when transition metals are adsorbed on the surfaces of 2D J-MoSSe. To our knowledge, few researchers have investigated the synergistic effects on the photocatalytic properties when vacancies *and* adsorbed transition metals coexist on the 2D J-MoSSe surface. Understanding synergistic effects in these materials is a valuable scientific question; the answer will provide insights into the photocatalytic activity of these imperfect 2D J-MoSSe surfaces. We also believe that exploring the interactions between different types of defects is an important scientific question for developing practical photolytic applications of co-doped semiconducting surfaces. In this work, we systematically study the geometric, electronic, optical absorption and surface adsorption properties of doped 2D J-MoSSe with intrinsic vacancies and extrinsic doping. Our research contributes to a comprehensive understanding for the synergistic effects of coexisting extrinsic doping and intrinsic vacancies, and our research also provides insights for the experimental preparation of doped 2D J-MoSSe structures for improving photocatalytic activity.

2. **Computational Methodology**

The 1H-phase $MoS_2$ has a hexagonal crystal lattice with point group of $D_{3d}$, in which two S atom layers sandwich the Mo atom layer. The $MoS_2$ structure will transform to the 2D J-MoSSe structure by completely replacing one of the two S atom layers in $MoS_2$ with Se atoms. Figure S1 shows the geometric structures of pristine 2D J-MoSSe and $MoS_2$. Ti adatom and vacancy may migrate on the surface of 2D J-MoSSe. The study of migration barriers presents how much energy the defects need to diffuse on the surface and determines whether the defects will diffuse readily or not. Synergistic effects arise when vacancies and Ti adatoms coexist in 2D J-MoSSe. The coexisting defects we considered in our calculations include various combinations like coexisting vacancy with adsorbed Ti atom on a *hollow* or a Mo-*top* site of the S/Se-side surface, respectively, as represented in Figure 2. The representative examples shown are: 1) the coexisting intrinsic S atomic vacancy with adsorbed Ti atom on the *hollow* site of S-side surface (Ti$_S^{ad\text{-}h}$V$_S$), in Figure 2(a); 2) the coexisting intrinsic S atomic vacancies with adsorbed Ti atom on the Mo-*top* site of Se-side surface (Ti$_{Se}^{ad\text{-}t}$V$_S$), in Figure 2(b); (3) the coexisting vacancy with substituted Ti atom of sulfur (Ti$_S$V$_S$), in Figure 2(c); (4) the coexisting vacancy with substituted Ti atom of selenium (Ti$_{Se}$V$_S$), in Figure 2(d). We also consider calculations on the coexisting Se atomic vacancy with Ti atom as shown in Figure S2. The energetics of the various structural configurations enable us to check which configuration would likely be favorable in an experimental preparation. We first explore the variation of energy corresponding to the distance between Ti adatom and S/Se vacancy in 2D J-MoSSe (in a 9×9×1 supercell) as shown in Figure S3.

Based on the results of the lowest energy configurations, we build simplified 4×4×1 supercells, as shown in Figure S5, to calculate the electronic properties with a higher precision in order to investigate the photocatalytic properties. Understanding the potentiality of 2D J-MoSSe defect structures in photocatalysis, we calculate the band structure, effective mass, and Gibbs free energy, optical absorption spectra, and electrostatic potentials. There are plenty of flat energy bands in the band structure of our large supercells because the band structures of a supercell get folded into the first Brillouin zone.[30] The unfolded band structure is necessary to recover an effective primitive cell band structure; thereby, representing a *unit cell* band structure containing defect levels.[31] The transport efficiency of charge carriers is dictated by the effective mass of this unfolded band structure. The reaction barrier of HER is determined by the binding strength of H atoms at the reaction sites and the change of Gibbs free energy (also denoted by hydrogen adsorption free energy). All these aforementioned properties are closely related to the photocatalytic activity and provide estimates on which configurations will improve photocatalytic performance.

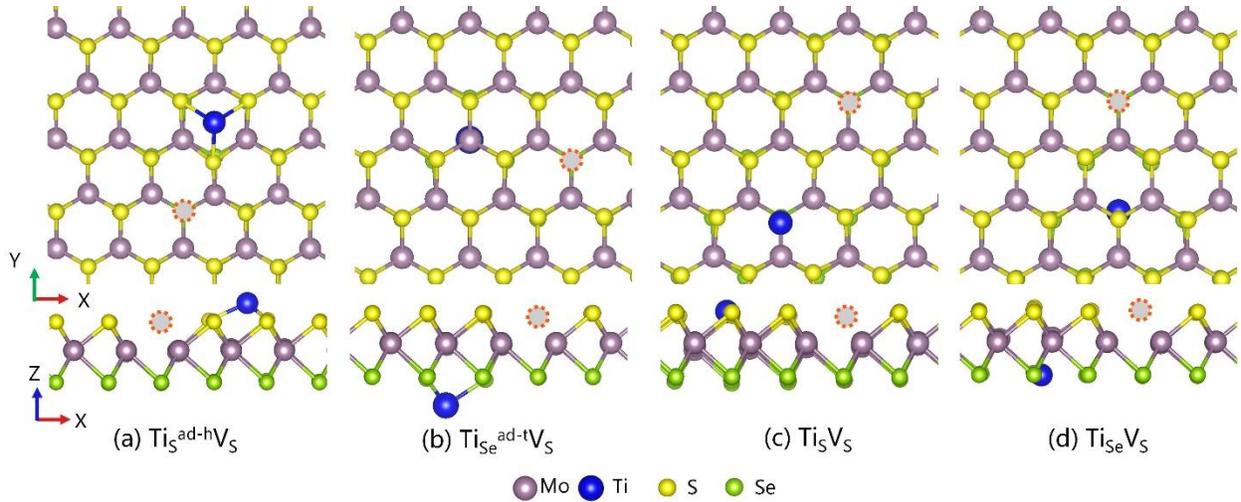

**Figure 2.** Schematic defect combinations in 2D J-MoSSe. (a) Top and side views of S vacancy coexisting with adsorbed Ti atom on the *hollow* site of the S-side surface ($Ti_S^{ad-h}V_S$). (b) Top and side views of S vacancy coexisting with adsorbed Ti atom on the *Mo-top* site of Se-side surface ($Ti_{Se}^{ad-t}V_S$). (c) Top and side views of S vacancy coexisting with substituted Ti atom of sulfur ($Ti_SV_S$). (d) Top and side views of S vacancy coexisting with substituted Ti of selenium ($Ti_{Se}V_S$). The dashed red circles indicate the location of S vacancies. The purple, blue, yellow and green balls represent the Mo, Ti, S and Se atoms, respectively.

In our calculations, we utilize the Quantum Espresso (QE)[32] and Vienna Ab initio Simulation Package (VASP)[33] software by means of the projected augmented wave (PAW) method.[34] The neb.x routine within the QE package helps to study the migration of S and Se vacancies and the adsorbed Ti adatoms via the nudged elastic band (NEB) method.[35] The unfolded band structures of the doped supercells clearly exhibit the effects of the defect states using the BandUP package.[36] We calculated the effective mass of charge carriers from the curvature of energy bands. The Atomic Simulation Environment (ASE),[37] written in Python, enables us to submit a large number of calculations. We utilized the thermochemistry modules in ASE to efficiently calculate the hydrogen adsorption free energy and the VASPKIT[38] tools for post-processing data generated by VASP.

## 3. Results and Discussion

The results in our work contains migration barrier, relative energy of different configurations and the photocatalysis related properties of configurations with lowest energy (optical absorption spectra, electrostatic potential difference, effective mass of charge carriers and hydrogen adsorption free energy). The configuration with Ti atom adjacent to the vacancy generally has the lowest energy. In these lowest energy configurations, adsorbed Ti atom dramatically extends the optical absorption range of 2D J-MoSSe, and the coexisting vacancy further increase the absorption coefficient. The Ti atoms on the Se-side surface greatly enhance the electrostatic potential difference. The Ti adatoms on matter existing alone or coexisting with vacancy will not greatly increase the effective mass of charge carriers and hydrogen adsorption free energies on these Ti-top sites are closer to zero compared to top site of substituted Ti atoms. These results indicate that the configurations with Ti adatom on the Se-surface exhibit mostly improved photocatalytic activities.

### 3.1 The migration of adsorbed Ti atoms and vacancies

We compare the energies of different configurations corresponding to Ti atoms adsorbed on different sites of the 2D J-MoSSe surface to find stable configurations. The Ti adatoms will possibly migrate from one *hollow* site to another *hollow* site on the surface of 2D J-MoSSe during the preparation process. The migration barriers for Ti adatoms are 0.8 eV on the S surface (0.6 eV on the Se surface) as shown in Fig. 2 (a). Interestingly, the *hollow* site ($Ti_S^{ad-h}$) is the most stable site for the Ti atom to be adsorbed on the S surface, while the Mo-*top* site ($Ti_{Se}^{ad-t}$) is the most stable site for the Ti atom to be adsorbed on Se surface. These results motivate us to only consider the two adatom cases in subsequent exploration: $Ti_S^{ad-h}$ and $Ti_{Se}^{ad-t}$.

The vacancies on the surface of 2D J-MoSSe may migrate similarly to Ti adatoms. The migration of S or Se vacancies is identical to the exchange of S or Se atom with the nearby vacancies, as shown in Fig. 1(b). We calculate the energy barriers as 1.7 eV or 1.5 eV for the migration of $V_S$ or $V_{Se}$, respectively. The vacancies will diffuse only with difficulty after appearing on the surface. Our results show that migration barriers on the S side are larger than corresponding values on Se side for both Ti adatoms and vacancies. We hypothesize that this is due to the larger electronegativity of sulfur leading to stronger bonds with Ti and Mo atoms compared with selenium.

### 3.2 The interaction between the coexisting Ti atoms and vacancies

As mentioned above, both the adsorbed Ti atoms and S/Se vacancies are difficult to migrate once these defects appear on the surface of 2D J-MoSSe. The adsorbed sites of Ti atoms may appear near or far from the vacancy site. The detailed variation of energy with respect to adatom/vacancy distances is valuable for us to develop better models for calculating photocatalytic properties. There would obviously be different energy scales when considering defects distributed on the two different S/Se surfaces. The combination of two defect sites weakens the lattice distortion and reduces the total energy. We systematically investigate the energy of various possible configurations of 2D J-MoSSe with different Ti adatom and vacancy locations. There are four

cases – 1) The Ti adatom at hollow site and vacancy exist on the S-surface (Ti$_S^{ad-h}$V$_S$); 2) Ti adatom at Mo-top site and vacancy on the Se-surface (Ti$_{Se}^{ad-t}$V$_{Se}$); 3) Ti at the hollow site on the S-surface and vacancy exist on the S-surface (Ti$_S^{ad-h}$V$_{Se}$); 4) Ti at the Mo-top site on the Se-surface and vacancy exist on the S-surface (Ti$_{Se}^{ad-t}$V$_S$); In this work, the distance is defined as the variation of Ti adatoms adsorbed at different sites with respect to a S/Se vacancy as Figure S3 shows.

In the simplest cases where the adatoms are located on the same surface as the vacancy, we find that the total energy decreases slightly as Ti adatoms get closer to a S/Se vacancy, as shown in Figure 3 (c). As the vacancy and Ti adatoms become adjacent, the energy dramatically decreases by approximately 1 eV (2 eV). These results indicate that the vacancy traps the adsorbed Ti atom nearby because there is an effective attraction between the two types of defects. In these cases, the Ti adatom migrates to the vacancy site and transforms to a substituted Ti atom, shown in Figure S3 (b).

For the defects on both surfaces, we first notice that the energy scale is dramatically different compared to the previous simplest cases. The total energy increases about 0.1 eV (0.3 eV) for configurations of Ti$_S^{ad-h}$V$_{Se}$ (Ti$_{Se}^{ad-t}$V$_S$) as the distance increases. The vacancy obviously cannot trap the Ti adatom because the molybdenum layer separates the two defects. The energy increase of Ti$_S^{ad-h}$V$_{Se}$ configuration (0.3 eV) is faster than that of Ti$_{Se}^{ad-t}$V$_S$ configuration (0.1 eV). A stronger bonding between the Ti adatom on the S surface leads to a stronger Coulomb attraction between the Ti adatom and the localized electron at the Se vacancy. This stronger Coulomb attraction induces a larger increase in total interaction energy compared to the pristine structure. The Ti$_{Se}^{ad-t}$V$_S$ configuration is weaker compared to the Ti$_S^{ad-h}$V$_{Se}$ because the Ti adatom binds more strongly to sulfur.

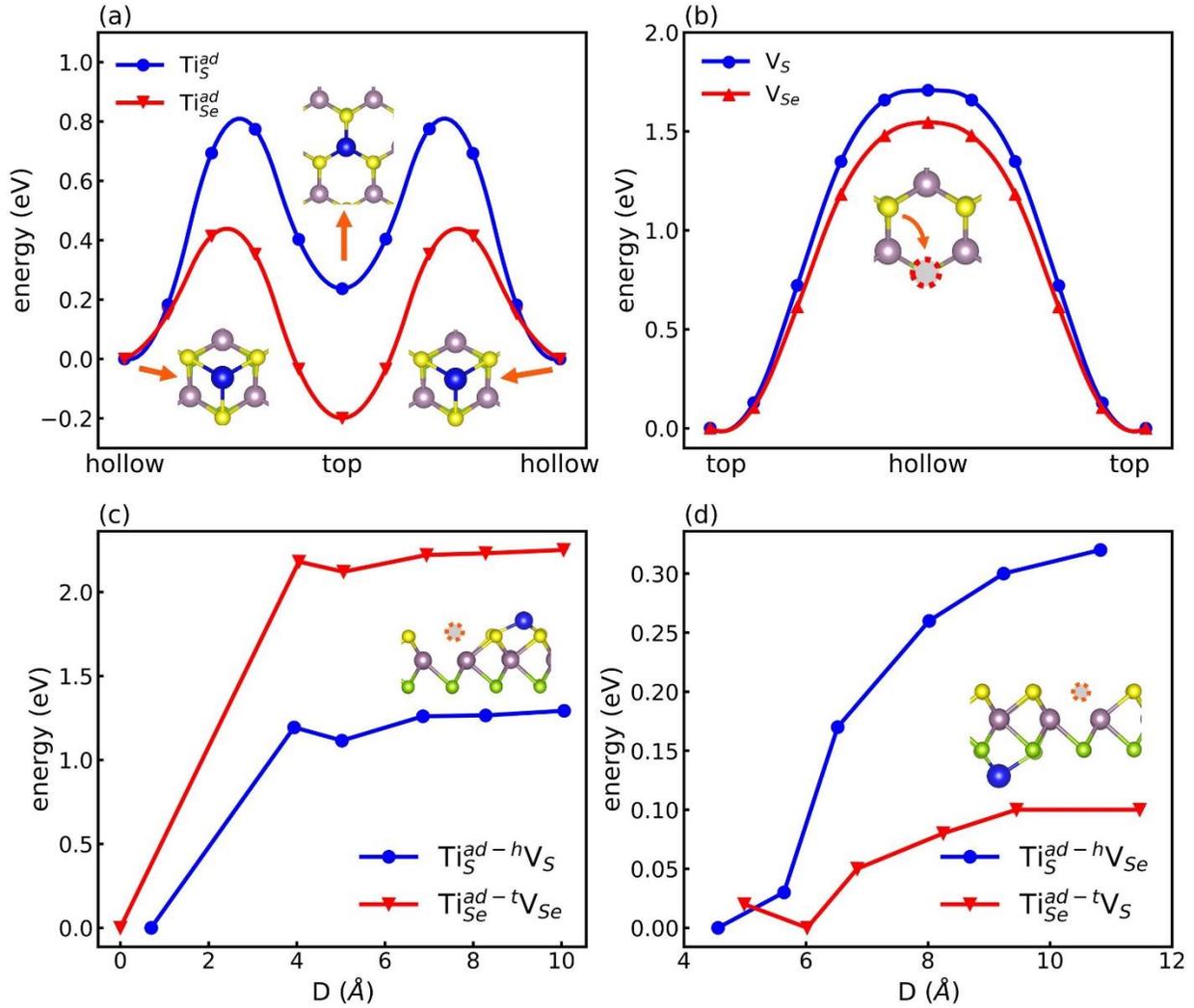

**Figure 3**. (a) The energy variation for adsorbed Ti atoms migrating from one *hollow* site to the adjacent *hollow* site by passing a middle Mo-top site. The energies are set to zero for adsorbed Ti atoms at the *hollow* site. (b) The energy variation for $V_S$ and $V_{Se}$ migrating from a top site to another top site by passing a *hollow* site. The energies are set to zero for S or Se atoms at the top site. (c) The relative energy of configurations varying with the distance (D) between Ti adatom coexisting and S/Se vacancies coexisting on one surface of 2D J-MoSSe. (d) The relative energy varying with distance between Ti adatom and S/Se vacancies on both surfaces. The lowest energy is set to zero for each type of configuration.

These results generally demonstrate that vacancies will trap nearby Ti adatoms. Both the lattice distortions and Coulomb interactions affect the total energy of these adatom-vacancy configurations. Therefore, the adsorbed Ti atoms are more likely to distribute at sites adjacent to vacancies or even get trapped by the vacancies when they are on same surface. A coexisting substituted Ti atom and vacancy on the 2D J-MoSSe surface may appear if there are excess vacancies (when the number of vacancies is greater than the number of Ti atoms). We find that the energy also decreases with decreasing distance between the substituted Ti atom and S/Se vacancy as shown in Figure S4; these results are similar to what we found for the Ti adatoms and S/Se vacancies. Ti atoms (substituted or adatom) coexisting on the sites adjacent to S/Se vacancies will produce configurations with the lowest energies. These lowest energy configurations are thereby thermodynamic favorable; we select these configurations to further our exploration of photocatalytic properties in the non-pristine 2D J-MoSSe system.

## 3.3 Optical Spectra

The utilization of sunlight is a critical key for energy conversion in photocatalysis. Defects of Ti atoms (substituted and adatoms) and S/Se vacancies will introduce defect states, thereby extending the optical absorption range of 2D J-MoSSe materials. The calculated absorption range of pristine 2D J-MoSSe includes the sunlight region above the photon energy of 1.7 eV with the first absorption peak at about 2.1 eV as shown in Figure 4(a). The 2D J-MoSSe structure exhibits improved sunlight utilization compared with $MoS_2$, which is consistent with the published results.[22] As a baseline for more complex configurations, we explore the properties of configurations with isolated Ti atoms or S/Se vacancies and used these results for comparison to the more complex configurations. All the configurations explored in this work are shown in Figure S5.

First, we find that vacancies, $V_S$ or $V_{Se}$, will extend the calculated absorption limit to 1.3 eV or 1.5 eV, respectively, as shown in Figure 4(a). The vacancies will also increase the absorption coefficients by filling valleys in the absorption spectra of pristine 2D J-MoSSe. Next, we find that the adsorbed Ti atoms on the Se-side surface dramatically extends the optical absorption limit by approximately 0.6 eV compared to the previously discussed substituted Ti coexisting with vacancies. This increased optical absorption will enable the absorption range to cover most of the visible region, as seen in Figure 4(b). We also explored the optical absorption spectra varying with different concentrations of adsorbed Ti atoms. And the results show that a 3%~8% ratio of adsorbed Ti atom has a remarkable enhancement and even a small doping ratio of 0.9% exhibits obvious improvement as described in Figure S6. Substituted Ti atoms ($Ti_S$ and $Ti_{Se}$) only slightly extend the calculated optical absorption range to about 1.6 eV compared with the pristine 2D J-MoSSe. The optical spectra of defects existing alone demonstrate that Ti adatoms induce broader optical absorption range than substituted Ti atoms and S/Se vacancies.

Of course, the S/Se vacancies are likely to coexist with *either* adsorbed or substituted Ti atoms on the surface of 2D J-MoSSe. So now, we will discuss these more complex situations. The coexisting S vacancy in the $Ti_{Se}^{ad-t}V_s$ configuration narrows the optical absorption range about 0.2 eV compared with the $Ti_{Se}^{ad-t}$ configuration as shown in Figure 4(b). The S vacancy further increases the absorption coefficient by filling valleys in the absorption spectra. The increased absorption coefficient mostly distributes at the energy range from 1.5 to 3 eV which is ideal for the photocatalytic water splitting because this energy range is in the region of higher-powered sunlight. The adsorbed Ti atom on the S-side surface ($Ti_S^{ad-t}$) has similar performance as shown in Figure

S6(c). For the configurations of vacancy coexisting with substituted Ti atoms ($Ti_SV_S$, $Ti_SV_{Se}$, $Ti_{Se}V_S$, and $Ti_{Se}V_{Se}$), we find that the coexisting S/Se vacancies further extend the calculated absorption limit to about 1.3 eV as shown in Figure S6 (a) and (b). The coexisting S/Se vacancies also increase the optical absorption coefficients. These spectra illustrate that the S/Se vacancies coexisting with Ti adatoms obviously increase the optical absorption coefficients and improve the conversion efficiency of solar energy. The configuration of S/Se vacancies coexisting with substituted Ti atoms have only a moderately-extended optical absorption range and increased absorption coefficient, which will also improve the conversion efficiency of solar energy.

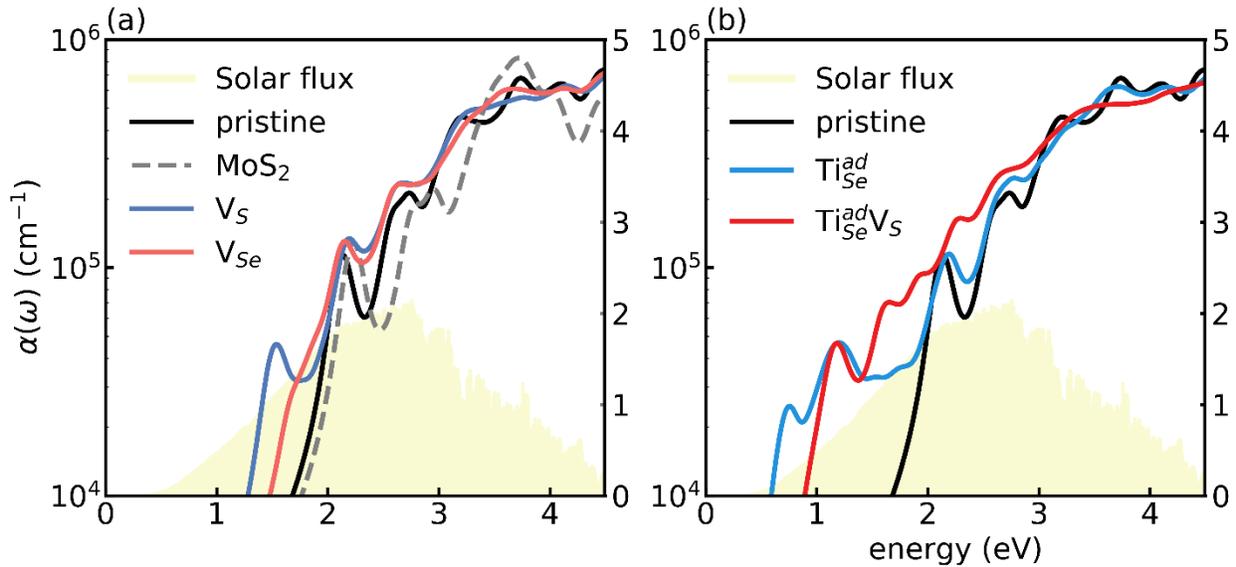

**Figure 4.** The optical absorption spectrum of pristine and defected 2D J-MoSSe. (a) the spectra of pristine 2D J-MoSSe, $MoS_2$ and MoSSe with S vacancy ($V_S$) and Se vacancy ($V_{Se}$). (b) illustrates the spectra of configurations containing adsorbed Ti atoms ($Ti_{Se}^{ad-t}$) and coexisting vacancies ($Ti_{Se}^{ad-t}V_S$). $Ti_{Se}^{ad}$ are the simplification of $Ti_{Se}^{ad-t}$ for convenience. The black curves in figure (a) and (b) are the spectrum of pristine MoSSe. The light-yellow shadows indicate the flux of solar radiation.[39]

Overall, the Ti adatoms and S/Se vacancies coexisting have the optimal synergistic effect for the harvest of sunlight when distributed on both surfaces of 2D J-MoSSe. We propose that the synergistic effects come from redistributed defect states when Ti adatom and S/Se vacancies coexist. The unfolded band structures of pristine 2D MoSSe, Ti$_{Se}^{ad-t}$, V$_S$, and Ti$_{Se}^{ad-t}$V$_S$ present clear figures of the defect states as shown in Figure S7. The S vacancy (V$_S$) by itself introduces shallower defect levels to the band structure as compared to Ti adatom (Ti$_{Se}^{ad-t}$) by itself, as shown in Figure S7 (b) and (d). Our calculations of Ti$_{Se}^{ad-t}$ configurations demonstrate a broader optical absorption range compared to the calculations of configurations with S vacancy (V$_S$) as shown in Figures S7(c). The first absorption peak at 0.6 eV corresponds to the transition of electrons from defect levels to the conduction band minimum (CBM).

In the more complex configurations (Ti adatoms coexisting with S vacancies, Ti$_{Se}^{ad-t}$V$_S$), our calculations demonstrate that the band structure exhibits more dispersive defect levels compared to the previously aforementioned simpler cases. The first absorption peak is located at 1.2 eV for Ti$_{Se}^{ad-t}$V$_S$ and this corresponds to the transition of electrons from valence band maximum (VBM) to the shallow defect levels as shown in Figure S7(e). The S vacancy within Ti$_{Se}^{ad-t}$V$_S$ configuration slightly narrows the optical absorption range compared to the Ti$_{Se}^{ad-t}$ configuration as shown in Figure S7 (c). A slightly blue-shifted first absorption peak in the spectrum of the Ti$_{Se}^{ad-t}$V$_S$ optical absorption calculation indicates that the deep defect states would not trap the photogenerated electrons, thereby discouraging the charge carriers from transporting to the potential level of the redox reaction. Our calculations indicate similar results for the Ti$_S^{ad-h}$V$_{Se}$ configuration as shown in Figure S6(c), but the optical absorption range is narrower than the absorption range of Ti$_{Se}^{ad-t}$V$_S$. These results demonstrates that the properties of the Ti$_{Se}^{ad-t}$V$_S$ configuration exhibit optimal synergistic effects within optical absorption spectrum which are more suitable for both harvesting sunlight and encouraging photocatalysis.

**3.4 The band edges with respect to the redox potentials of water**

The band edges of a photocatalyst should straddle the redox potentials of water splitting to enable the transition of electron-hole pairs from the conduction/valence bands to the redox potential levels. The pristine 2D J-MoSSe has a band gap of 2.04 eV as per our calculated results using the HSE06 functional, and our calculated CBM and VBM levels properly straddle the redox potentials as shown in Figure 1(a). For pristine 2D J-MoSSe, the calculated electrostatic potential difference (ΔV) is 0.77 eV consistent with reported results[24] which will drive the photo-generated electrons and holes to distribute on the Se and S surface respectively. Therefore, without defects, the band edges of pristine 2D J-MoSSe fulfills the requirements for photocatalysis. But the CBM is too close to the reductive potential without being increased by ΔV. With the electrostatic potential difference, ΔV, added, the CBM and VBM levels bend upwards along the direction of the internal electric field and the ΔV upshifts the CBM level and its density of states is located primarily on the Se side. The energy difference between CBM and the reductive potential increases and is comparable with the difference between the VBM and the oxidative potential level. The comparable energy difference will induce an analogous reaction rate between hydrogen evolution reaction (HER) and oxygen evolution reaction (OER) and improve the overall efficiency.

The defects of S/Se vacancies and Ti atoms (substituted or adsorbed) will enhance or reduce the ΔV by different extents. Our calculations show that the intrinsic S or Se vacancies hardly change the ΔV compared to the pristine 2D J-MoSSe. Both adsorbed and substituted Ti atoms will increase the ΔV when distributing on the Se surface, while decrease ΔV when on the S surface. The Ti adatoms in the Ti$_{Se}^{ad-t}$ configuration increase the ΔV to 1.66 eV, while the Ti adatoms in Ti$_S^{ad-h}$ configuration decrease the ΔV to 0.11 eV. These changes result from Ti ions existing as centers of positive charge as the outmost electrons of Ti atoms will transfer to S or Se atoms. These Ti ions with positive charge will enhance the internal electric field of MoSSe pointing from Se to S atom layers when the ions distribute on Se-surface, while weaken the electric field when Ti ions distribute on the S-surface, which is consistent with reported results.[21] The Ti adatom introduces a larger increase to ΔV than the substituted one. These results indicate that either adsorbed or substituted Ti atoms on the Se-side surface are more favorable to enhance the ΔV of 2D J-MoSSe.

The doped Ti atoms along with existing S/Se vacancies on the surface of 2D J-MoSSe will also adjust ΔV and will shift the band edges. The S vacancy tends to slightly increase ΔV, while the Se vacancy decreases ΔV a bit. The defects in Ti$_{Se}^{ad-t}$V$_S$ configuration induce a slightly larger ΔV of 1.71 compared to Ti$_{Se}^{ad-t}$. The results of ΔV indicate that only Ti atoms (substituted or adsorbed) on the S-side surface decrease the ΔV no matter if the Ti atoms exist alone or coexist with S/Se vacancies. There are no obvious changes for the CBM and VBM levels of configuration V$_S$, Ti$_{Se}^{ad-t}$V$_S$ and Ti$_{Se}^{ad-t}$ compared to the pristine 2D J-MoSSe as shown in Figure S7. The band edges of these configurations still distribute on the Se and S surface respectively and CBM levels bend upwards to straddle the redox potential levels of water splitting reaction similarly to the pristine 2D J-MoSSe.

**Table 1.** The electrostatic potential difference (ΔV) between the two surfaces for pristine and defected 2D J-MoSSe. The value of -0.17 eV indicates the reversed ΔV with direction contrary to the intrinsic potential difference in pristine 2D J-MoSSe.

| Configuration | Pristine | $V_S$ | $Ti_S^{ad-h}$ | $Ti_S^{ad-h}V_{Se}$ | $Ti_S$ | $Ti_S V_S$ | $Ti_S V_{Se}$ |
|---|---|---|---|---|---|---|---|
| ΔV(eV) | 0.77 | 0.77 | 0.11 | -0.17 | 0.24 | 0.26 | 0.19 |
| Configuration | | $V_{Se}$ | $Ti_{Se}^{ad-t}$ | $Ti_{Se}^{ad-t}V_S$ | $Ti_{Se}$ | $Ti_{Se}V_{Se}$ | $Ti_{Se}V_S$ |
| ΔV(eV) | | 0.73 | 1.66 | 1.71 | 1.11 | 1.05 | 1.12 |

### 3.5 The carrier separation and transport

The separation and transport of photo-induced carriers are another key step for the photocatalytic water splitting which greatly affect the efficiency of the redox reaction. The internal electric field not only shifts the band edges, but also boosts the separation of electrons and holes in 2D J-MoSSe. The intrinsic electrostatic potential difference of 2D J-MoSSe will effectively boost the separation of charge carriers.[40] The charge carriers will separate more efficiently in the configurations with Ti atom (adsorbed or substituted) on the Se-surface as the increased ΔV regardless the Ti atom is existing alone or coexisting with vacancies. The $Ti_{Se}^{ad-t}$ and $Ti_{Se}^{ad-t}V_S$ configurations will possess a higher efficiency for photocatalytic water splitting because the internal electric field is enhanced by the change in electrostatic potential difference, ΔV.

The effective mass dictates the transport efficiency of charge carriers. We study the effective mass from the curvature of the conduction/valence energy bands based on the unfolded band structures as Figure S8 shows. From the carrier's effective mass shown in Figure 5, the pristine 2D J-MoSSe has effective carrier mass smaller than the free electron mass $m_0$ and exhibits very weak x-y direction anisotropy and hole-electron asymmetry. For the defects existing alone, the vacancies ($V_S$ and $V_{Se}$) have minor effects on the carrier effective mass. The Ti adatoms ($Ti_S^{ad-h}$ and $Ti_{Se}^{ad-t}$) also have little impact on the effective mass. The substituted Ti atoms ($Ti_S$ and $Ti_{Se}$) severely increase the effective mass both of electrons and holes to about 2~ $3m_0$. The defect $Ti_{Se}$ introduce distinct differences between the effective mass of electrons and holes which will promote the separation of hole-electron pairs. The effective mass results of $V_{Se}$ and $Ti_S^{ad-h}$ are shown in Figure S9. These effective mass results demonstrate that only substituted Ti atoms induce large effective mass of charge carriers and yield decreases in the transport efficiency of charge carriers. The S/Se vacancies and Ti adatoms induce small changes in the effective mass of charge carriers, thereby having little impact on the efficient transporting of charge carriers in 2D J-MoSSe.

More factors contribute to the changes in the effective mass for systems where S/Se vacancies and Ti atoms (substituted or adsorbed) coexist leading to complex interpretations. The S vacancy and Ti adatoms in $Ti_{Se}^{ad-t}V_S$ configuration induce increases in the carrier mass. The effective mass of electrons slightly increases to about $m_0$ and exhibit a large x-y direction anisotropy for $Ti_{Se}^{ad-t}V_S$ configuration. The defects in the $Ti_S^{ad-h}V_{Se}$ configuration yield larger effective masses for electrons compared to holes as shown in Figure S9, thereby promoting separation of electron-hole pairs. Substituted Ti atoms coexisting with S/Se vacancies in the $Ti_SV_S$, $Ti_{Se}V_{Se}$, $Ti_SV_{Se}$ and $Ti_{Se}V_S$ configurations have obviously negative impact on the effective mass of charge carriers. The larger effective mass of $Ti_{Se}V_S$ configuration exhibit an apparent x-y direction anisotropy compared to the Ti adatom configurations ($Ti_{Se}^{ad-t}V_S$ and $Ti_S^{ad-h}V_{Se}$) because the substituted Ti atom coexisting with S vacancy introduces a larger lattice distortion into the initial structure, thereby breaking the local symmetry of the 2D J-MoSSe structure. The broken local symmetry reduces the curvature and enhances the x-y direction anisotropy of conduction/valence bands as shown in Figure S8 (f).

The defects containing Ti adatoms with S/Se vacancies only slightly increase the effective mass of charge carriers, while the defects containing substituted Ti atoms increase the effective mass and reduce the transport efficiency of photo-generated electrons and holes.

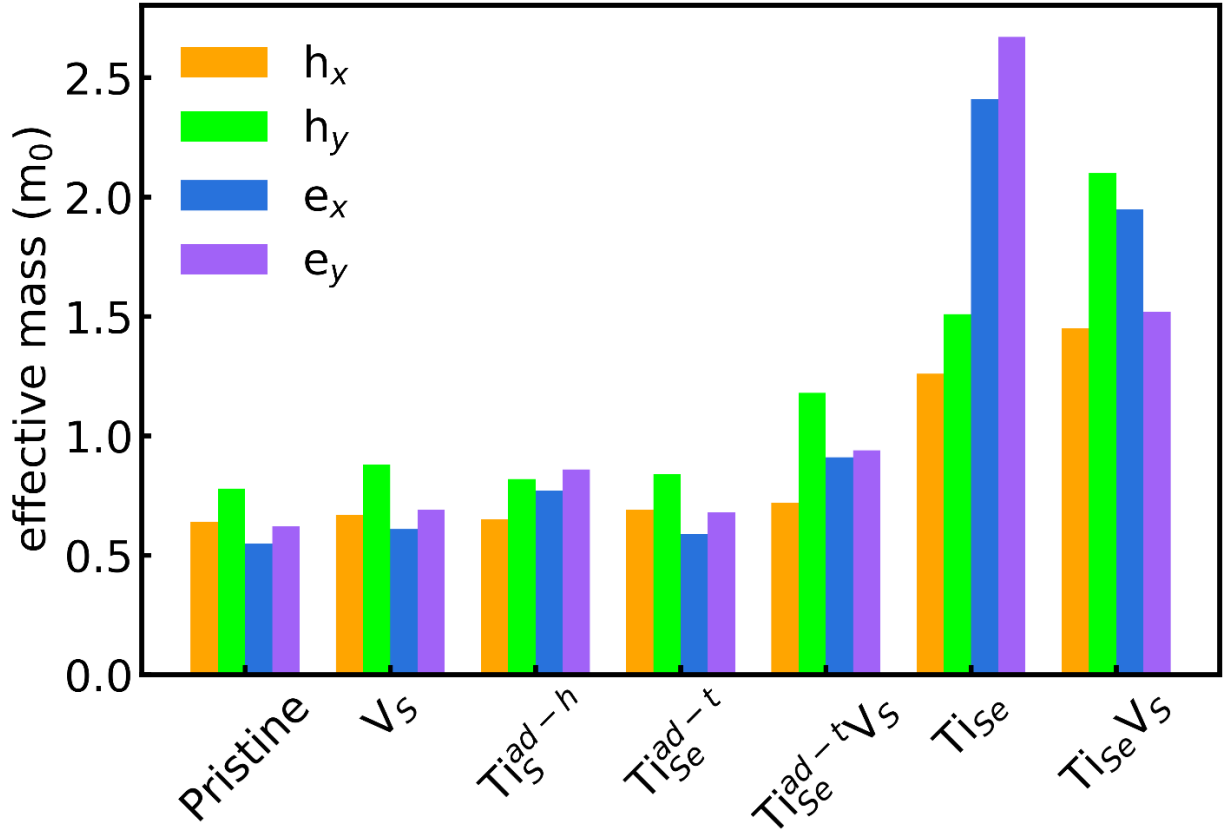

**Figure 5.** The effective mass of electron and hole in pristine and defected 2D J-MoSSe. The $h_x$, $h_y$, $e_x$ and $e_y$ denote the effective mass of holes and electrons in the x and y direction, respectively.

### 3.6 The surface catalytic reaction

The hydrogen evolution reaction (HER) is the last step for the hydrogen generation in photocatalytic water splitting. The reaction barrier for reducing the hydrogen ion is determined by the free energy ($\Delta G_{H*}$) and characterizes the reaction rate for HER. There are multiple active sites in doped 2D J-MoSSe for the adsorption of hydrogen atom including S/Se vacancy sites and Ti-top sites. The $\Delta G_{H*}$ barriers of pristine 2D J-MoSSe are too high yielding very poor performance to catalyze HER as shown in Figure S10. For isolated defects, the S/Se vacancies in the $V_S$ and $V_{Se}$ configurations decrease the barrier $\Delta G_{H*}$ to -0.1 eV and 0.01 eV, respectively, and these $\Delta G_{H*}$ values imply excellent catalytic activity which is consistent with the experiment results.[20] The Ti adatom in $Ti_S^{ad-h}$ configuration also induces near zero $\Delta G_{H*}$ values similar to the configurations containing only S/Se vacancies. The Ti adatoms in the $Ti_{Se}^{ad-t}$ configuration induce a $\Delta G_{H*}$ of approximately -0.2 eV as shown in Figure 6, which is still appropriate for catalyzing HER.[26, 41, 42] In contrast, substitutional Ti atoms within the $Ti_S$ and $Ti_{Se}$ configurations induce larger $\Delta G_{H*}$ values of approximately 0.9 eV as shown in Figure S10.

Now we consider the very complex configurations containing both S/Se vacancies and Ti atoms (either substitutional or adatoms). The defects in $Ti_{Se}^{ad-t}V_S$ configuration also induce moderate changes in the $\Delta G_{H*}$ barriers compared to the Ti-top site of the $Ti_{Se}^{ad-t}$ configuration without vacancy. The $\Delta G_{H*}$ values of vacancy sites within the $Ti_SV_S$, $Ti_{Se}V_S$ and $Ti_SV_{Se}$ configurations are approximately -0.1 eV as shown in Figure S10. These vacancy sites are also HER active. From the free energy values discussed above, the vacancy and Ti adatom top sites are generally the most catalytically active sites for HER. While the substituted Ti atoms will not improve the HER catalytic efficiency.

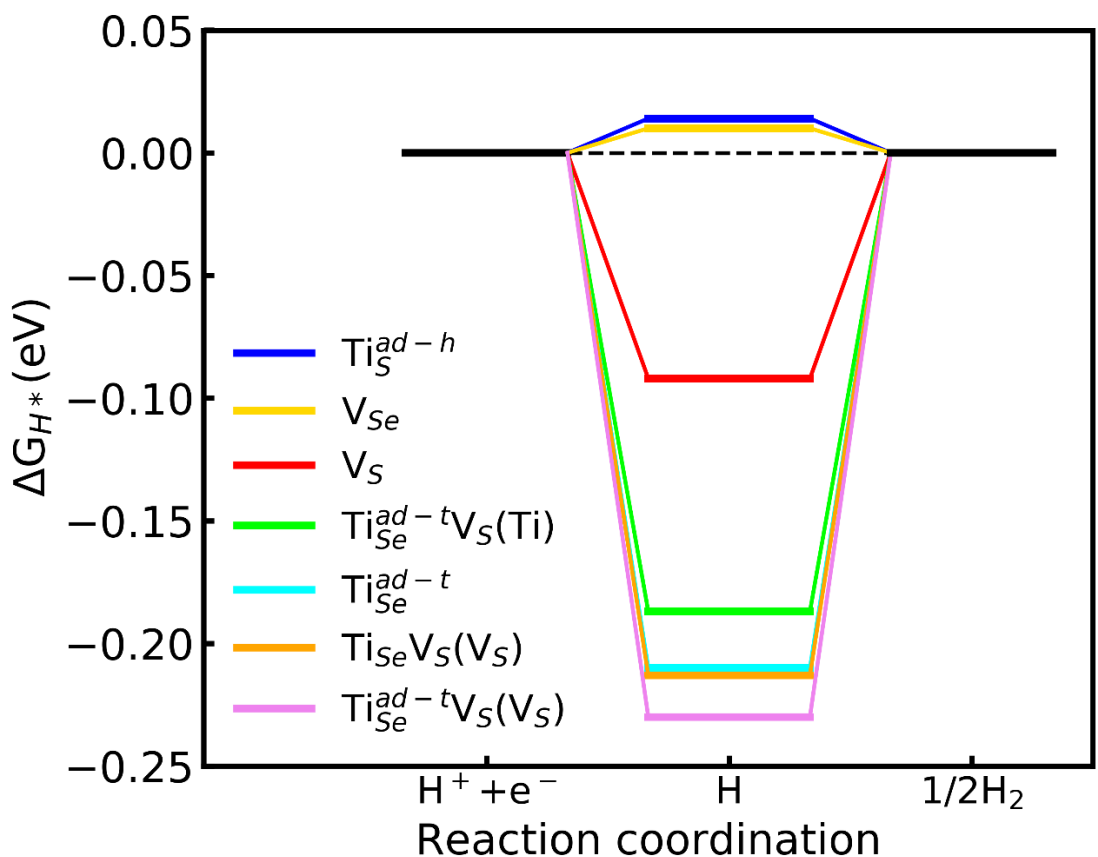

**Figure 6.** The change of Gibbs free energy ($\Delta G_{H*}$) for HER on vacancy or Ti-top sites of defected 2D J-MoSSe. $Ti_{Se}^{ad-t}V_S$ (Ti) means the Ti-top site, and $Ti_{Se}^{ad-t}V_S$ ($V_S$) means the vacancy site in configuration $Ti_{Se}^{ad-t}V_S$. The sites are HER active with $\Delta G_{H*}$ between -0.2 and 0.2 eV.

## 4   Conclusion

The properties of configurations above should meet as many requirements mentioned above as possible for photocatalysis with a series of steps. The rules we take for selecting optimal configurations are as follows: First, the defects should not weaken the internal electric field greatly; Second, these defects should improve the optical absorption properties by either extending the optical absorption range or increasing the absorption coefficient; Third, the structure should contain active sites for HER; Last, the charge carriers should separate efficiently and transfer quickly to the reaction sites. Although Ti adatoms in the Ti$_S^{ad-h}$ configuration extend the optical absorption range and decrease $\Delta G_{H^*}$ values very close to zero, the Ti atom reduces the $\Delta V$ from 0.77 eV to 0.1 eV, which diminishes the main advantage of 2D J-MoSSe. So, the Ti adatoms on the Se surface is superior. The coexisting defects in Ti$_{Se}^{ad-t}V_S$ configuration increase the $\Delta V$ to 1.7 eV, extend the optical absorption limit of 0.9 eV, increase the absorption coefficient, and decrease the hydrogen adsorption free energy barrier to -0.19 eV. We predict that this Ti$_{Se}^{ad-t}V_S$ configuration exhibits the best overall photocatalytic performance among all the configurations we had considered. The Ti$_S^{ad-h}$ and V$_{Se}$ configurations also look promising based on similar properties aforementioned. Substituted Ti atoms show less likelihood photocatalytic activity, and we propose that these systems should be avoided. If one hopes to obtain improved photocatalytic applications, then one should also prepare materials with few vacancies, particularly, the Se surface. However, Ti adatoms deposited on the Se surface will transform to the Ti$_{Se}^{ad-t}V_S$ configuration if there are S vacancies which will produce a promising photocatalytic material.

In conclusion, we have demonstrated by systematically exploring the structures and properties of 2D J-MoSSe (with Ti doping and S/Se vacancies) that there are synergistic effects on the photocatalytic activity. Adsorbed Ti atoms will migrate on the surface during the preparation process, while S or Se vacancies cannot migrate given the higher energy barrier. The doped Ti atoms energetically prefer sites adjacent to vacancies. Adsorbed Ti atoms enlarge the range of optical absorption dramatically, and coexisting vacancies will further increase the absorption coefficient. Overall, Ti atoms will strengthen the internal electric field only when adsorbing on the Se side. The vacancy site and top site of adsorbed Ti atom are generally HER catalytic active with hydrogen adsorption free energies in the range from -0.2 eV to 0.2 eV. The Ti$_{Se}^{ad-t}$V$_S$ configuration does exhibit better overall photocatalytic activity by taking the above factors into account.

ASSOCIATED CONTENT

**Supporting Information**.

The following files are available free of charge.

Supplementary Figures of the structure of MoS$_2$ and MoSSe, Structure of doped 2D Janus MoSSe, definition of the distance, variation of energy with distance, configurations explored in detail, optical absorption spectra, unfolded band structures, effective mass, hydrogen adsorption energy and free energy. (PDF)

AUTHOR INFORMATION

**Corresponding Author**

*E-mail: james.p.lewis.phd@gmail.com

*E-mail: yan@ucas.ac.cn

*E-mail: gsu@ucas.ac.cn

**ORCID**

James P. Lewis: 0000-0002-6724-3483

Qing-Bo Yan: 0000-0002-1001-1390

Gang Su: 0000-0002-8149-4342

**Notes**

The authors declare no competing financial interest.

ACKNOWLEDGMENT

This work is supported in part by the National Key R&D Program of China (Grant No. 2018YFA0305800), the Strategic Priority Research Program of CAS (Grant No. XDB28000000), the NSFC (Grant No. 11834014), Beijing Municipal Science and Technology Commission (Grant No. Z118100004218001), JPL received support from the Thousand Talent program of the Chinese Academy of Sciences and Fundamental Research Funds for the Central Universities. The authors thank Prof Zheng-Chuan Wang, Prof Zhen-Gang Zhu, Prof Bo Gu, Yurong He, Zhongxian Qiu, and Kuan-Rong Hao for valuable discussion.

ABBREVIATIONS

2D J-MoSSe, two dimensional Janus MoSSe; CVD, Chemical Vapor Deposition; CBM, Conduction Band Minimum; VBM, Valence Band Maximum; HER, Hydrogen Evolution Reaction; OER, Oxygen Evolution Reaction.